\documentclass[twocolumn,aps,pra,groupedaddress,showpacs]{revtex4}
\usepackage[english]{babel}
\usepackage{float}
\usepackage{color}
\usepackage{amsmath}
\usepackage{graphicx}
\usepackage{amssymb}
\usepackage{txfonts}
\usepackage[unicode=true,pdfusetitle,
 bookmarks=false,bookmarksnumbered=false,bookmarksopen=false,
 breaklinks=false,pdfborder={0 0 1},backref=false,colorlinks=false]
 {hyperref}

 \definecolor{BLACK}{gray}{0}
 \definecolor{WHITE}{gray}{1}
 \definecolor{RED}{rgb}{1,0,0}
 \definecolor{GREEN}{rgb}{0,1,0}
 \definecolor{BLUE}{rgb}{0,0,1}
 \definecolor{CYAN}{cmyk}{1,0,0,0}
 \definecolor{MAGENTA}{cmyk}{0,1,0,0}
 \definecolor{YELLOW}{cmyk}{0,0,1,0}

\makeatother

\begin{document}

\title{Necessary and sufficient condition for quantum adiabatic evolution by unitary control fields}

\author{Zhen-Yu Wang}
\thanks{Part of this work was carried out at the Department of Physics, The Chinese University of Hong Kong, Shatin,
N. T., Hong Kong.}
\affiliation{Institut f\"ur Theoretische Physik, Albert-Einstein-Allee 11, Universit\"at Ulm, 89069 Ulm, Germany}
\author{Martin B. Plenio}
\affiliation{Institut f\"ur Theoretische Physik, Albert-Einstein-Allee 11, Universit\"at Ulm, 89069 Ulm, Germany}

\begin{abstract}
We decompose the quantum adiabatic evolution as the products of gauge invariant unitary operators and obtain the exact nonadiabatic correction in the adiabatic approximation. A necessary and sufficient condition that leads to adiabatic evolution with geometric phases is provided and we determine that in the adiabatic evolution, while the eigenstates are slowly varying, the eigenenergies and degeneracy of the Hamiltonian can change rapidly. We exemplify this result by the example of the adiabatic evolution driven by parametrized pulse sequences. For driving fields that are rotating slowly with the same average energy and evolution path, fast modulation fields can have smaller nonadiabatic errors than obtained under the traditional approach with a constant amplitude.
\end{abstract}


\maketitle

\hyphenpenalty 1800
\section{Introduction}
The adiabatic theorem in quantum mechanics concerns the evolution
of quantum systems subject to slowly varying Hamiltonians~\cite{Messiah:1965:North}.
It states that the transitions between the instantaneous eigenstates
of a Hamiltonian are negligible if the change of the Hamiltonian is
much slower than the energy gaps between the instantaneous eigenstates.
Berry discovered that in addition to the dynamic phase the 
adiabatic evolution exhibits a geometric phase determined 
only by the path~\cite{Berry:1984:45}.
Wilczek and Zee generalized the result to non-Abelian geometric phase
for degenerate Hamiltonians~\cite{Wilczek:1984:2111}. While the
adiabatic theorem has a wide range of applications, it was found that
the widely used adiabatic quantitative condition \begin{equation}
\left|\frac{\langle n_{p}^{t}|\dot{m}_{q}^{t}\rangle}{E_{n}(t)-E_{m}(t)}\right|\ll1,\label{eq:ADC1}
\end{equation}
for adiabatic approximation can be invalid~\cite{Marzlin:2004:160408,Tong:2005:110407,Sarandy:2004:331,Du:2008:060403,Amin:2009:220401}. Here $|n_{p}^{t}\rangle$ ($|m_{q}^{t}\rangle$) is the Hamiltonian eignenstate with the eigenenergy $E_{n}(t)$ [$E_{m}(t)$] and degeneracy label $p$ ($q$), and the dot means a time derivative.
As a consequence of these observations a debate arose and new adiabatic conditions were proposed
(e.g., Refs.~\cite{Tong:98:150402,Wei:2007:024304,Comparat:2009:012106,lidar:2009:102106,Boixo:2010:032308,Rigolin2012Adiabatic,Guo2013Nonperturbative}).
Those works~\cite{Marzlin:2004:160408,Tong:2005:110407,Sarandy:2004:331,Du:2008:060403,Amin:2009:220401,Tong:98:150402,Wei:2007:024304,Comparat:2009:012106,lidar:2009:102106,Boixo:2010:032308,Rigolin2012Adiabatic,Guo2013Nonperturbative}
and the debate on the necessity of Eq.~(\ref{eq:ADC1})~\cite{Tong:2010:120401,Zhao2011Comment,Comparat2011Comment,Tong2011Tong,Li:2014:053023}),
however, start from the assumption of non-degenerate Hamiltonians
with a gap condition (i.e., $|E_{n}(t)-E_{m}(t)|>0$). It has been noted however, that the formulation of an adiabatic theorem with finite
numbers of energy crossings is possible~\cite{Avron:1999:203}. To
verify the adiabatic conditions in the general setting, it is important
to obtain the exact nonadiabatic correction in the adiabatic approximation
for Hamiltonians with possible energy crossings.

In this work, we consider Hamiltonians $H(t)$ with possible energy
degeneracies and arbitrary numbers of energy crossings. We decompose the quantum
evolution
\begin{equation}
U(t)=U_{\text{Dyn}}(t)U_{\text{Geo}}(t)U_{\text{Dia}}(t),\label{eq:U:Ud-Ugeo-Uerr}
\end{equation}
as the products of unitary operators: the dynamic
phase operator $U_{\text{Dyn}}(t)$, the geometric phase operator
$U_{\text{Geo}}(t)$, and the nonadiabatic correction $U_{\text{Dia}}(t)$
in the adiabatic approximation. In the adiabatic limit, $U_{\text{Dia}}(t)=I$
is the identity operator and $U_{\text{adia}}(t)=U_{\text{Dyn}}(t)U_{\text{Geo}}(t)$
is the exact adiabatic evolution. From $U_{\text{\text{Dia}}}(t)$,
we derive an upper bound of the nonadiabatic deviation in the adiabatic
approximation and propose a necessary and sufficient condition for
adiabatic evolution. Counterintuitively perhaps, we find that the eigenenergies
of the Hamiltonian can change rapidly and can have an arbitrary number
of energy crossings during the adiabatic evolution. The result presented here reveals
that the crucial condition for adiabatic evolution is a slowly varying eigenpath, while the eigenenergies
are not required to vary slowly. This finding leads to a new
way to realize adiabatic evolution. By applying a sequence of coherent
pulses or a fast varying field parameterized by the adiabatic path,
we can achieve the adiabatic evolution with accumulated (non-Abelian)
geometric phases in a shorter time for a given average energy.

\section{Gauge invariant formalism for adiabatic evolution}
Here we obtain the exact nonadiabatic deviation and derive the general condition for adiabatic evolution.
Consider a quantum system driven by a time-dependent Hamiltonian $H(t)\equiv H(\boldsymbol{R})\equiv H(\vartheta)$,
where $\boldsymbol{R}\equiv\left(R_{1}(\vartheta),R_{2}(\vartheta),\cdots\right)$
is parametrized by the dimensionless parameter $\vartheta=\vartheta(t)$. And 
\begin{equation}
\omega=\omega(t)\equiv \frac{d\vartheta}{dt},
\end{equation}
describes the speed of traversing a path. The function parameters
$t$, $\boldsymbol{R}$, and $\vartheta$ are used interchangeably
in this paper. The evolution of arbitrary quantum states from the
moment $t=0$ (with the parameters $\boldsymbol{R}=\boldsymbol{R}_{0}$
and $\vartheta=\vartheta_{0}$) to the moment $T$ (i.e., $\boldsymbol{R}_{T}$
and $\vartheta_{T}$) is described by the evolution operator $U(T)$,
which satisfies the Schr\"odinger equation ($\hbar=1$)
\begin{equation}
i\dot{U}(t)=H(t)U(t).\label{eq:SchrodingerEq}
\end{equation}

The instantaneous orthonormal eigenstates $|n_{j}^{\boldsymbol{R}}\rangle\equiv|n_{j}^{\vartheta}\rangle\equiv|n_{j}^{t}\rangle$
at the moment $t$ satisfy
$H(t)|n_{j}^{t}\rangle=E_{n}(t)|n_{j}^{t}\rangle\equiv E_{n}(\vartheta)|n_{j}^{\vartheta}\rangle$.
Substituting
the transformation $U(t)\equiv U_{1}(t)U_{2}(t)$ in Eq.~(\ref{eq:SchrodingerEq})
with $U_{1}(t)$ a unitary operator, we obtain $i\dot{U}_{2}(t)=H_{2}(t)U_{2}(t)$ with $H_{2}(t)=U_{1}^{\dagger}(t)\left[H(t)-i\dot{U}_{1}(t)U_{1}^{\dagger}(t)\right]U_{1}(t)$
in the interaction picture~\cite{Feynman:1951:108}. By another transformation
$U_{1}(t)=U_{\text{Dyn}}(t)U_{\text{G1}}(t)$
with
\begin{align}
U_{\text{Dyn}}(t)& \equiv \sum_{n,j}e^{-i\int_{0}^{t}E_{n}(t^{\prime})dt^{\prime}}|n_{j}^{t}\rangle\langle n_{j}^{t}|,\label{eq:Udyn}\\
U_{\text{G1}}(t)& \equiv\sum_{n,j}|n_{j}^{t}\rangle\langle n_{j}^{0}|,\label{eq:UG1}
\end{align}
we obtain
$H_{2}(t)=-i\sum_{n\neq m;p,q}|n_{p}^{0}\rangle\langle n_{p}^{t}|\dot{m}_{q}^{t}\rangle\langle m_{q}^{0}|e^{i\int_{0}^{t}(E_{n}-E_{m})dt^{\prime}}+H_{\text{G2}}(t)$ with $H_{\text{G2}}(t)=-i\sum_{n,p,q}|n_{p}^{0}\rangle\langle n_{p}^{t}|\dot{n}_{q}^{t}\rangle\langle n_{q}^{0}|$, where $|n_{j}^{0}\rangle$ are the initial eigenstates.
To obtain the nonadiabatic correction $U_{\text{Dia}}(t)$, we write $U_{2}(t)$ as
$U_{2}(t)=U_{\text{G2}}(t)U_{\text{Dia}}(t)$, where
\begin{equation}
U_{\text{G2}}(t)\equiv\mathcal{T}\exp\left[-i\int_{0}^{t}H_{\text{G2}}(t^{\prime})dt^{\prime}\right],\label{eq:UG2}
\end{equation}
with $\mathcal{T}$ the time ordering operator. In the decomposition
\begin{equation}
U(t)= U_{\text{Dyn}}(t)U_{\text{G1}}(t)U_{\text{G2}}(t)U_{\text{Dia}}(t), \label{eq:UDynG1G2Dia}
\end{equation}
$U_{\text{Dyn}}(t)$ is the dynamic phase
operator and $U_{\text{Geo}}(t)\equiv U_{\text{G1}}(t)U_{\text{G2}}(t)$ is the geometric phase
operator. The geometric phase operator
\begin{equation}
U_{\text{Geo}}(t)=\sum_{n,j}|n_{j}^{\boldsymbol{R}}\rangle\langle n_{j}^{\boldsymbol{R}_{0}}|\mathcal{P}e^{i\int_{\boldsymbol{R}_{0}}^{\boldsymbol{R}}\sum_{n,p,q}|n_{p}^{\boldsymbol{R}_{0}}\rangle\langle n_{p}^{\boldsymbol{R}^{\prime}}|i\nabla_{\boldsymbol{R}^{\prime}}|n_{q}^{\boldsymbol{R}^{\prime}}\rangle\langle n_{q}^{\boldsymbol{R}_{0}}|\cdot d\boldsymbol{R}^{\prime}},\label{eq:UGeo}
\end{equation}
is generally non-Abelian for degenerate Hamiltonians. Here
$\mathcal{P}$ is the path ordering operator on $\boldsymbol{R}$
or $\vartheta$, and $\nabla_{\boldsymbol{R}}\equiv(\frac{\partial}{\partial R_{1}},\frac{\partial}{\partial R_{2}},\cdots)$
acts on $|n_{q}^{\boldsymbol{R}}\rangle$. The nonadiabatic correction reads
\begin{equation}
U_{\text{Dia}}(t)=\mathcal{P}\exp\left[i\int_{\vartheta_{0}}^{\vartheta}\sum_{n\neq m;p,q}F_{n,m}(\vartheta^{\prime})G_{n,m}^{p,q}(\vartheta^{\prime})d\vartheta^{\prime}\right],\label{eq:Herr}
\end{equation}
where the geometric functions
\begin{equation}
G_{n,m}^{p,q}(\vartheta)\equiv U_{\text{Geo}}^{\dagger}(\vartheta)|n_{p}^{\vartheta}\rangle\left(\langle n_{p}^{\vartheta}|i\frac{d}{d\vartheta}|m_{q}^{\vartheta}\rangle\right)\langle m_{q}^{\vartheta}|U_{\text{Geo}}(\vartheta),\label{eq:Hnmpq}
\end{equation}
describe nonadiabatic transitions $|m_{q}^{\vartheta}\rangle\leftrightarrow|n_{p}^{\vartheta}\rangle$,
and the modulation functions
\begin{equation}
F_{n,m}(\vartheta)\equiv e^{i\int_{0}^{t}[E_{n}(t^{\prime})-E_{m}(t^{\prime})]dt^{\prime}}=e^{i\int_{\vartheta_{0}}^{\vartheta}[E_{n}(\vartheta^{\prime})-E_{m}(\vartheta^{\prime})]\frac{1}{\omega}d\vartheta^{\prime}},\label{eq:filterFunc}
\end{equation}
are determined by the energy gaps $E_{n}(t)-E_{m}(t)$ and the speed of
path sweeping $\omega$. We have separated the effects of $F_{n,m}$
(determined by the eigenenergies $E_{n}$) and $G_{n,m}^{p,q}$ (determined
by eigenstates $|n_{j}^{t}\rangle$) in $U_{\text{Dia}}(t)$. The
decomposition Eq.~(\ref{eq:U:Ud-Ugeo-Uerr}) is obtained, with $U_{\text{Dia}}(t)$ describing all the nonadiabatic effects.

An important property of our general formalism
is that the unitary operators $U_{\text{Dyn}}(t)$, $U_{\text{Geo}}(t)$, and $U_{\text{Dia}}(t)$
are all gauge invariant (see Appendix~\ref{sec:GaugeInv}). That is, these unitary operators
do not change when we replace $|n_{j}^{t}\rangle$ with $W_{t}|n_{j}^{t}\rangle$
in the formulas, where $W_{t}$ is a time-dependent unitary transformation
of degenerate subspaces with the property $\langle m_{p}^{t}|W_{t}|n_{q}^{t}\rangle=0$
if $m\neq n$. An example of $W_{t}$ is the phase-shift operator
of the eigenstates, $W_{t}=\sum_{n,j}e^{i\phi_{n,j}(t)}|n_{j}^{t}\rangle\langle n_{j}^{t}|$. Examples of gauge invariant operators are the system Hamiltonian $H(t)$ and the corresponding unitary propagator $U(t)$. Examples of operators that are not gauge invariant are $U_{\text{G1}}(t)$ [Eq.~(\ref{eq:UG1})] and  $U_{\text{G2}}(t)$ [Eq.~(\ref{eq:UG2})]. Not all decompositions of a gauge invariant unitary operators are gauge invariant. For example, $U_a(t)=U_{\text{Dyn}}(t)U_{\text{G1}}(t)$ and $U_b(t)=U_{\text{G2}}(t)U_{\text{Dia}}(t)$, for a different decomposition $U(t)=U_a(t)U_b(t)$ of Eq.~(\ref{eq:UDynG1G2Dia}), are not gauge invariant.

\section{Condition for quantum adiabatic evolution}

The deviation from the adiabatic evolution is described by
\begin{equation}
D_{\text{Dia}}(t)\equiv U_{\text{Dia}}(t)-I.\label{eq:Derr}
\end{equation}
When its
unitarily invariant norm~\cite{Lidar:2013:QEC,comment:operatorNorm} $\left\Vert D_{\text{Dia}}(t)\right\Vert \approx0$,
the quantum evolution is adiabatic with $U_{\text{Dia}}(t)\approx I$.

Let the average of the modulation functions be bounded by $\xi_{\text{avg}}$ during the evolution time $T$,
\begin{equation}
\left|\int_{\vartheta_{0}}^{\vartheta}F_{n,m}(\vartheta^{\prime})d\vartheta^{\prime}\right|<\xi_{\text{avg}},\forall \vartheta\in[\vartheta_{0},\vartheta_{T}]\text{ and }n\neq m.\label{eq:averagingCondition}
\end{equation}
Note that the left-hand side of Eq.~(\ref{eq:averagingCondition}) is the absolute value of the Fourier component 
\begin{equation}
f_{n,m}(\lambda)\equiv\int_{\vartheta_{0}}^{\vartheta}F_{n,m}(\vartheta^{\prime})e^{-i\lambda\vartheta^{\prime}}d\vartheta^{\prime},
\end{equation}
at $\lambda=0$.

If $\int_{\vartheta_{j}}^{\vartheta_{j+1}}F_{n,m}(\vartheta)d\vartheta=0$ for the intervals $\vartheta_{j+1}-\vartheta_{j}<\eta$ with $j=0,1,\ldots, N$ and $\vartheta_{N+1}\equiv\vartheta_{T}$, we have $\xi_{\text{avg}}=\eta$. For this partition, the upper bound of the nonadiabatic correction reads (see Appendix~\ref{sec:Condition})
\begin{equation}\left\Vert D_{\text{Dia}}(T)\right\Vert <\xi_{\text{avg}}\left(g_{\text{tot}}^{2}+w_{\text{tot}}\right)(\vartheta_{T}-\vartheta_{0}),
\label{eq:DBound}
\end{equation}
where $g_{\text{tot}}=\sum_{n\neq m}g_{n,m}$ and $w_{\text{tot}}=\sum_{n\neq m}w_{n,m}$
with the least upper bounds $g_{n,m}\equiv\sup_{\vartheta\in[\vartheta_{0},\vartheta_{T}]}||\sum_{p,q}G_{n,m}^{p,q}(\vartheta)||$
and $w_{n,m}\equiv\sup_{\vartheta\in[\vartheta_{0},\vartheta_{T}]}||\sum_{p,q}\frac{d}{d\vartheta}G_{n,m}^{p,q}(\vartheta)||$. Note that the factor $\left(g_{\text{tot}}^{2}+w_{\text{tot}}\right)(\vartheta_{T}-\vartheta_{0})$ on the right-hand side of Eq.~(\ref{eq:DBound}) is only a function of evolution path.

In Appendix~\ref{sec:Condition}, we also derive two upper bounds in general cases for $\xi_{\text{avg}}\ll 1$, i.e., 
\begin{equation}
\left\Vert D_{\text{Dia}}(t)\right\Vert \lesssim 2(\vartheta-\vartheta_{0})\sqrt{\xi_{\text{avg}}g_{\text{tot}}\left(g_{\text{tot}}^{2}+w_{\text{tot}}\right)}, \label{eq:bound:type1}
\end{equation}
and
\begin{align}
\left\Vert D_{\text{Dia}}(t)\right\Vert < &  \sqrt{\xi_{\text{avg}}}\left(g_{\text{tot}}^{2}+w_{\text{tot}}\right)(\vartheta-\vartheta_{0})^2 \nonumber \\
& + (\sqrt{\xi_{\text{avg}}}+\xi_{\text{avg}}) g_{\text{tot}}. \label{eq:bound:type2}
\end{align}

To be valid for arbitrary finite smooth paths, the averaging condition
(\ref{eq:averagingCondition}) with vanishing $\xi_{\text{avg}}\rightarrow 0$
can be shown to be necessary and sufficient for the adiabatic approximation
$U_{\text{Dia}}(t)\rightarrow I$ during $t\in[0,T]$ (see Appendix~\ref{sec:Condition}). The sufficiency is obvious from the bounds Eqs.~(\ref{eq:bound:type1}) or (\ref{eq:bound:type2}), and the physical reason is the following. The condition (\ref{eq:averagingCondition}) means
that the low-frequency Fourier components $f_{n,m}(\lambda)$
of $F_{n,m}(\vartheta^{\prime})$ are negligible when $\xi_{\text{avg}}\ll1$, since for a small $\lambda$
the factor $e^{-i\lambda\vartheta^{\prime}}$ is slowly varying and we can show
$f_{n,m}(\lambda)\approx0$ by the generalized Riemann-Lebesgue lemma~\cite{Kahane:1980:108,Li:2008:229}.
The condition $\xi_{\text{avg}}\rightarrow 0$ is sufficient because $F_{n,m}(\vartheta)$
are fast oscillating functions and the slowly varying functions $G_{n,m}^{p,q}(\vartheta)$
are averaged out. If the adiabatic limit $U_{\text{Dia}}(t)\rightarrow I$
is valid for arbitrary finite smooth paths, we can always find some
paths which lead to $\xi_{\text{avg}}\rightarrow 0$ in Eq.~(\ref{eq:averagingCondition}), and thus
Eq.~(\ref{eq:averagingCondition}) with $\xi_{\text{avg}}\rightarrow 0$ is also necessary.

To have a simple picture of the condition Eq.~(\ref{eq:averagingCondition}), consider as an example the case that the ratios $r_{n,m}\equiv\frac{\hbar\omega}{E_{n}-E_{m}}$ of the speed $\omega$ for traversing a path to the energy gaps $E_{n}-E_{m}$ are constants. 
By using Eqs.~(\ref{eq:filterFunc}) and (\ref{eq:averagingCondition}), 
we obtain $|f_{n,m}(0)|\leq 2|r_{n,m}|$. Therefore, we can choose $\xi_{\text{avg}}=\text{max} |r_{n,m}|+0^{+}$ for the condition Eq.~(\ref{eq:averagingCondition}). For finite energy gaps $E_{n}-E_{m}$, the slow evolution limit $\omega=d\vartheta/dt\rightarrow 0$ (i.e., the limit of infinite evolution time $t \rightarrow \infty$) gives $\xi_{\text{avg}}\rightarrow 0$ and hence the quantum adiabatic evolution. Note that since the time and energy are conjugate variables, we can realize the quantum adiabatic evolution by increasing the energy gaps $|E_{n}-E_{m}|\rightarrow \infty$ for a finite speed $\omega$ and finite evolution times $t$. If we treat $\hbar\omega$ as the energy scale of the excitation caused by path transversal, we have another physical interpretation. The adiabatic approximation is valid when the excitation energy scale $\hbar\omega$ is much smaller than the energy gap $E_{n}-E_{m}$. 

The above arguments apply to situations that the energy gaps and the speed $\omega$ change smoothly, since we can split the evolution into pieces and the evolution of each piece has approximately constant ratios $r_{n,m}$.

\section{Adiabatic evolution by pulse sequences}
\begin{figure}
\includegraphics[width=0.5\columnwidth]{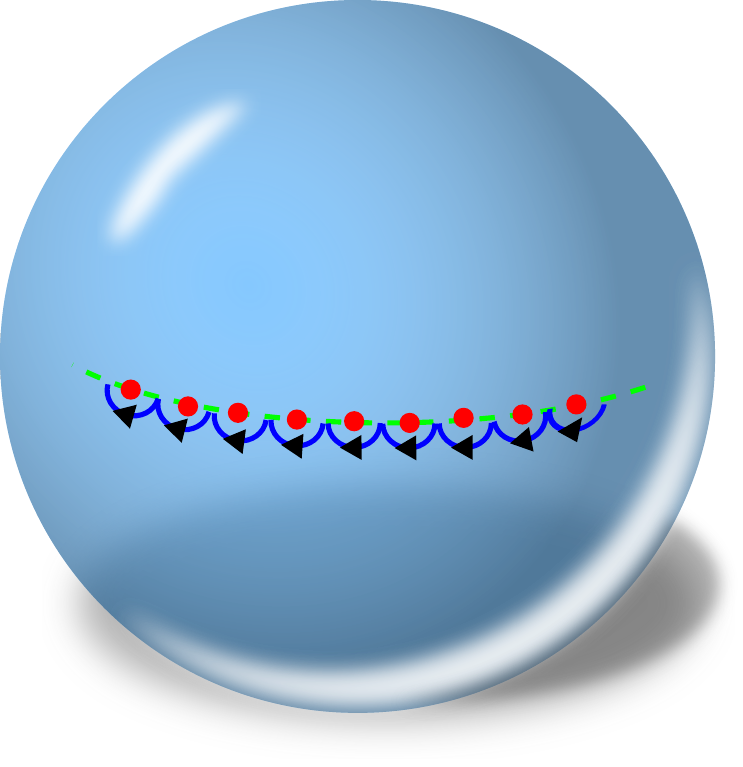}\caption{\label{fig:Path}(Color online). Illustration of a sequence of pulses applied to a
spin-$\frac{1}{2}$. The red dots indicate the directions of the pulses
on the Bloch sphere. The blue semi-circles illustrate the directions
of spin rotation. The green dashed line shows the effective path of
adiabatic evolution. These successive rotations induce a geometric
phase.}
\end{figure}
Now we show that adiabatic evolution can be driven by pulsed Hamiltonians. We consider a quantum system driven solely by a sequence of $N$ unitary pulses
\begin{equation}
P(\boldsymbol{R}_{k})=\sum_{n,j}e^{-i\theta_{n}(\boldsymbol{R}_{k})}|n_{j}^{\boldsymbol{R}_{k}}\rangle\langle n_{j}^{\boldsymbol{R}_{k}}|.\label{eq:UnitaryPulses}
\end{equation}
The idea is illustrated by a two-level system in Fig.~\ref{fig:Path}.
Between the pulses there is no control and the system is gapless with $H(t)=0$~\cite{comment:H0}, which is not the setting of previous  works~\cite{Marzlin:2004:160408,Tong:2005:110407,Amin:2009:220401,Sarandy:2004:331,Tong:98:150402,Wei:2007:024304,Comparat:2009:012106,lidar:2009:102106,Boixo:2010:032308,Tong:2010:120401}.
The pulses are applied in the order of the
parameters $\boldsymbol{R}_{k}=\boldsymbol{R}_{1}$, $\boldsymbol{R}_{2}$,
$\ldots,$ $\boldsymbol{R}_{N}$, which sample a path gradually, and they induce the modulation functions $F_{n,m}(\vartheta)$ to average out the effects of nonadiabatic transitions. The actual time duration of each pulse can be arbitrary (within the coherence time).
For $M$ non-degenerate subspaces, we can choose $\theta_{n}(\boldsymbol{R}_{k})=2\pi n/M$
with $n=1,\cdots,M$. If the system is a spin-$J$ system, the pulses are just rotations with an angle $2\pi/(2J+1)$ by a magnetic field that defines
the eigenstates $|n_{j}^{\boldsymbol{R}_{k}}\rangle$. If we apply
the pulses equidistantly during the parameter range $[\vartheta_{0},\vartheta_{T}]$,
the integral $\int_{\vartheta_{0}}^{\vartheta}F_{n,m}(\vartheta^{\prime})d\vartheta^{\prime}=O(1/N)$
vanishes at large $N$.
The dynamic phase is $\sum_{k}\theta_{n}(\boldsymbol{R}_{k})$ and the
geometric phase factor $U_{\text{Geo}}(T)$ is given by Eq.~(\ref{eq:UGeo})
with the path sampled by the points $\boldsymbol{R}_{k}$.

Note that this pulse sequence is different from dynamical decoupling pulse sequences~\cite{Viola:1998:2733,BanJMO1998,Yang:2010:2}, which also use pulses to induce modulation functions to average out unwanted evolutions~\cite{Wang:2013:042319}. Here the pulses are parametrized by a path sampled by $\{\boldsymbol{R}_{k}\}$
and are used to suppress state transitions caused by the change of system eigenstates, whereas dynamical decoupling uses pulses to suppress unwanted terms in the original Hamiltonians (e.g., interactions from environments). Additionally, to suppress unwanted terms, the control Hamiltonian used in dynamical decoupling generally does not commute with the original system Hamiltonian. 
For example, in using dynamical decoupling to suppress the pure dephasing (caused by the noise along the $z$ direction) of a qubit, the control fields are required to have components perpendicular to $z$ (the control fields along the $z$ direction cannot suppress the noise along the $z$ direction). 
In contrast, the Hamiltonian to generate the pulse sequences for adiabatic evolution is the total Hamiltonian (in the rotating frame of the bare Hamiltonian).

Another way to traverse an adiabatic path is the use of a sequence
of projective measurements~\cite{john1996mathematical,Childs:2002:032314}, which
can be simulated by evolution randomization~\cite{Boixo:2009:833,Chiang:2014:012314}.
If we begin in the ground state of $H(\boldsymbol{R}_{0})$ and successively
measure $H(\boldsymbol{R}_{1}),H(\boldsymbol{R}_{2}),\cdots,H(\boldsymbol{R}_{N})$,
then the final state will be the ground state of $H(\boldsymbol{R}_{N})$
with high probabilities, assuming that the difference between successive
points is sufficiently small. 
Unlike those methods~\cite{john1996mathematical,Childs:2002:032314,Boixo:2009:833,Chiang:2014:012314}, 
our protocol represents
a deterministic quantum algorithm to stroboscopically sample an intended path and is easier to implement in experiments.

\begin{figure}
\includegraphics[width=1\columnwidth]{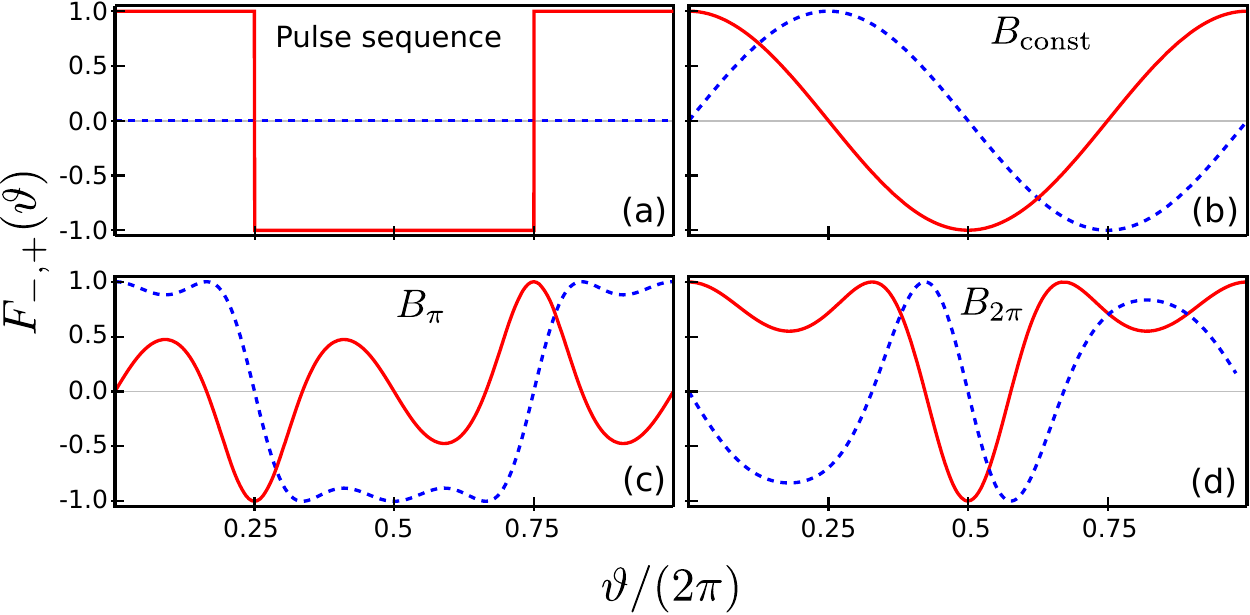}\caption{\label{fig:Filter}(Color online). The modulation functions $F_{-,+}(\vartheta)$
of a spin-$\frac{1}{2}$ in a scaled period. Red solid (blue dashed)
lines for the real (imaginary) part of $F_{-,+}(\vartheta)$.
(a) for the modulation function of a pulse sequence, (b) for a constant
field $B_{\text{const}}$, and (c) and (d) for fast varying fields
$B_{\pi}$ and $B_{2\pi}$, respectively.}
\end{figure}

\subsection{A spin-$\frac{1}{2}$ driven by a pulse sequence}
An example of the pulses in Eq.~(\ref{eq:UnitaryPulses}) for a spin-$\frac{1}{2}$
is a sequence of equidistant $\pm\pi$ rotations along the directions
$\hat{\boldsymbol{x}}\sin\theta\cos\vartheta_{k}+\hat{\boldsymbol{y}}\sin\theta\sin\vartheta_{k}+\hat{\boldsymbol{z}}\cos\theta$
with
\begin{equation}
\vartheta_{k}=(\vartheta_{T}-\vartheta_{0})\left(\frac{2k-1}{2N}\right)+\vartheta_{0},\text{ for \ensuremath{k}=1,\ensuremath{\ldots},}N,
\end{equation}
and $\vartheta_{0}=0$ (see Fig.~\ref{fig:Path}). Since the sampling
of $\vartheta$ is similar to the timing of Carr-Purcell (CP) sequences~\cite{Carr:1954:630},
we denote our sequence as $\text{CP}_{\text{Geo}}$ pulse sequence
for convenience. Each of the unitary pulse, $P(\vartheta_{k})=\sum_{\pm}\exp\left[\pm i(-1)^{s_{k}}\frac{\pi}{2}\right]|\vartheta_{k}^{\pm}\rangle\langle\vartheta_{k}^{\pm}|$
with $s_{k}\in\{\pm1\}$, introduces a $\pm\pi$ phase shift between
the instantaneous eigenstates $|\vartheta_{k}^{\pm}\rangle$. 
To isolate the geometric phase by cancelling the dynamic phase~\cite{Leek:2007:1889,Berger:2012:220502}, we can use equal numbers of $+\pi$ and
$-\pi$ pulses or even numbers of $\pi$ pulses.
The geometric (Berry) phase from $\vartheta_{0}$
to $\vartheta_{T}$ is $U_{\text{Geo}}(T)=\sum_{\pm}|\vartheta_{T}^{\pm}\rangle\langle\vartheta_{0}^{\pm}|e^{\pm i\frac{1}{2}\vartheta_{T}\cos\theta}$,
and
$U_{\text{Dia}}(T)=\mathcal{P}\exp\left[{\frac{-i}{2}\int_{\vartheta_{0}}^{\vartheta_{T}}\left(\sin\theta F_{-,+}(\vartheta)e^{i\cos\theta\vartheta}|\vartheta_{0}^{-}\rangle\langle\vartheta_{0}^{+}|+\text{H.c.}\right)d\vartheta}\right]$,
where the modulation function $F_{-,+}(\vartheta)=(-1)^{k}$
when $\vartheta\in(\vartheta_{k-1},\vartheta_{k}]$ {[}see Fig.~\ref{fig:Filter}(a){]}.
Note that if we apply $2\pi$ rotations on the spin-$\frac{1}{2}$,
even though the energy gaps are larger during the control, the modulation
function $F_{-,+}(\vartheta)=1$ does not have averaging
effects and the adiabatic evolution is not realized.

\begin{figure}
\includegraphics[width=1\columnwidth]{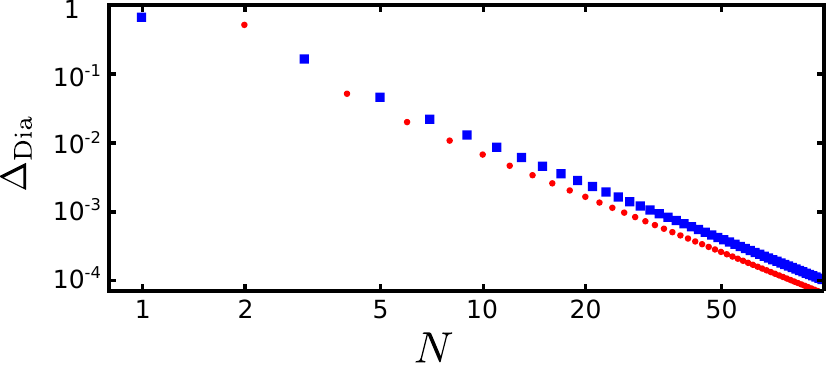}\caption{\label{fig:IdealCPMG}(Color online). The average deviation $\Delta_{\text{Dia}}$
as a function of $\text{CP}_{\text{Geo}}$ pulse number $N$, with
red circles (blue squares) for even (odd) $N$. Here $\vartheta_{0}=0$,
$\vartheta_{T}=2\pi$, and $\theta=\pi/6$. }
\end{figure}

We measure the nonadiabatic correction at the moment $T$ numerically
by the average deviation $\Delta_{\text{Dia}}\equiv\overline{|\langle\Psi|D_{\text{Dia}}(T)|\Psi\rangle|}$,
where the over bar is the average over all possible states
$|\Psi\rangle$. We plot the deviation $\Delta_{\text{Dia}}$ under
the control of $\text{CP}_{\text{Geo}}$ pulses in Fig.~\ref{fig:IdealCPMG},
which shows that as the pulse number increases, the nonadiabatic
evolution is smaller because of better averaging. The $\text{CP}_{\text{Geo}}$
sequences with even pulse numbers have better performance than
those with odd $N$. Note that when $\theta=\pi/2$, $D_{\text{Dia}}(T)=0$ for the $\text{CP}_{\text{Geo}}$
sequences with any pulse numbers $N\geq1$.

\subsection{A spin-$\frac{1}{2}$ driven by continuously varying fields \label{subsec:continous}}
Fast varying fields that are changing continuously can also lead to
adiabatic evolution and can have better performance than slowly varying
fields in traditional adiabatic evolution. Consider the driving fields
$B(t)\left(\hat{\boldsymbol{x}}\sin\theta\cos\vartheta+\hat{\boldsymbol{y}}\sin\theta\sin\vartheta+\hat{\boldsymbol{z}}\cos\theta\right)$
on a spin-$\frac{1}{2}$ with $\vartheta=\omega t$, where $B(t)$ has the values (i) $B_{\pi}(t)=\frac{\Omega}{2}[1+\gamma\cos(\Omega t)]$,
(ii) $B_{2\pi}(t)=2B_{\pi}(t)$, and (iii) $B_{\text{const}}(t)=\sqrt{(2+\gamma^{2})/8}\Omega$,
which has the same average energy as $B_{\pi}(t)$ {[}i.e., $\int_{0}^{\frac{2\pi}{\Omega}}|B_{\text{const}}|^{2}dt=\int_{0}^{\frac{2\pi}{\Omega}}|B_{\pi}|^{2}dt${]}.
We set $\gamma\approx2.34$ so that the average of the modulation
function $e^{i\int_{0}^{t}B_{\pi}(s)ds}$ vanishes in a half period
$\pi/\Omega$ (see Fig.~\ref{fig:Filter}). The eigenenergies are $\pm\frac{1}{2}B(t)$. There are degeneracy points for
$B_{\pi(2\pi)}(t)=0$. The field $B_{\pi(2\pi)}$ contributes a
$\pi$ ($2\pi$) phase shift in each period of $2\pi/\Omega$.

\begin{figure}
\includegraphics[width=1\columnwidth]{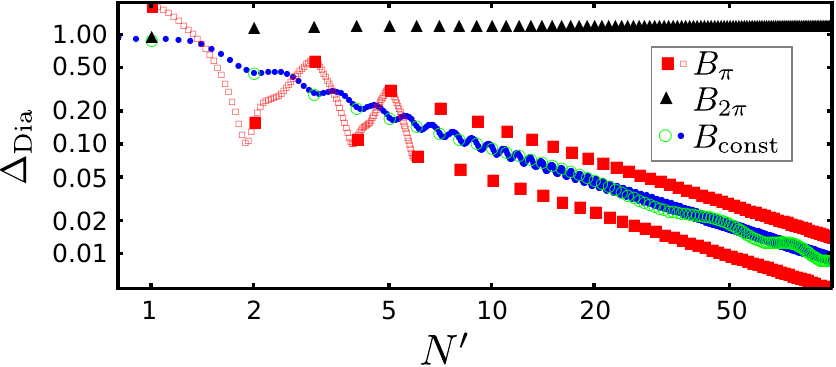} \caption{\label{fig:Sin}(Color online). The plot of $\Delta_{\text{Dia}}$
as a function of $N^{\prime}$ for the continuous driving with
$\vartheta_{0}=0$, $\vartheta_{T}=2\pi$, and $\theta=\pi/2$. The integer values of $N^{\prime}$ correspond to the numbers of applied $\pi$
pulses for the field $B_{\pi}$. The results are shown at integer numbers of $N^{\prime}$,
with the red squares (black triangles) for the fast varying field $B_{\pi}$
($B_{2\pi}$) and the green circles for the field $B_{\text{const}}$
of constant amplitude. For other values of $N^{\prime}$, the results
are shown for $B_{\text{const}}$ by blue dots and for $B_{\pi}$
by red empty squares from $N^{\prime}=1$ to $6$.}
\end{figure}

In Fig.~\ref{fig:Sin}, we plot $\Delta_{\text{Dia}}$ for $B_{\pi}$,
$B_{2\pi}$, and $B_{\text{const}}$ as a function of $N^{\prime}\equiv\Omega T/2\pi$
with $\omega T=2\pi$ and the total evolution time $T=1$. For $B_{\pi}$,
the integer values of $N^{\prime}$ are the numbers of accumulated $\pi$
phases during the evolution. Increasing $N^{\prime}$ (i.e., increasing
the energy) is equivalent to increasing the evolution time in adiabatic
evolution. As shown in Fig.~\ref{fig:Sin}, the fast varying field
$B_{\pi}$ realizes the adiabatic evolution even though the field
amplitude changes rapidly and there are many energy crossings during
the evolution. The field $B_{\pi}$ with even numbers of $\pi$ phase
shifts is much more efficient than the slowly varying field $B_{\text{const}}$
in traditional adiabatic evolution, because the modulation function
$e^{i\int B_{\pi}dt}$ is more efficient than $e^{i\int B_{\text{const}}dt}$
(see Fig.~\ref{fig:Filter}). Even though
$B_{2\pi}(t)$ has a larger amplitude and energy than $B_{\pi}(t)$,
it can not realize adiabatic evolution because the average of the
modulation does not vanish. Thus larger field amplitudes do not always
lead to better adiabatic evolution.

Note that here the energy crossings are not avoided crossings. With
perturbation, multiple avoided crossings can occur, and the effect
of multiple Landau-Zener transitions~\cite{Shevchenko:2010:1} is
a topic for future study.

\section{The Marzlin-Sanders inconsistency in degenerate Hamiltonians}
The quantitative condition Eq.~(\ref{eq:ADC1})
had been widely used as a criterion for the adiabatic approximation.
Unlike the condition in Eq.~(\ref{eq:averagingCondition}), the condition in Eq.~(\ref{eq:ADC1})
is a function of eigenstates (i.e., the evolution path) in
addition to the dependency on eigenenergies. The path
dependency may cause failure of adiabatic approximation for some evolution
paths. 

Indeed, it was first discovered by Marzlin and Sanders that
this condition~(\ref{eq:ADC1}) is not sufficient for adiabatic approximation~\cite{Marzlin:2004:160408,Tong:2005:110407}.
If a system $A$ with the Hamiltonian $H(t)$ follows the adiabatic
evolution and $|\langle n^{0}|n^{t}\rangle|\neq1$, another system
$\bar{A}$ driven by the Hamiltonian $\bar{H}(t)=-U^{\dagger}(t)H(t)U(t)$ with
$U(t)=\mathcal{T}e^{-i\int_{0}^{t}H(s)ds}$ cannot have adiabatic
evolution even if both systems satisfy the same condition (\ref{eq:ADC1}).
Here the overbar denotes quantities for the system $\bar{A}$.
The inconsistency for non-degenerate Hamiltonians was explained by
the resonant transitions between the energy levels in $\bar{H}(t)$~\cite{Amin:2009:220401}.

Here we consider general Hamiltonians with possible degeneracy and
show that the unbounded path of the second system $\bar{A}$ violates the adiabatic
approximation. It can be shown that the eigenstates of the system $\bar{A}$ are expressed by
the first system $A$ as 
\begin{equation}
|\bar{n}_{j}^{t}\rangle=U^{\dagger}(t)|n_{j}^{t}\rangle,\label{eq:njBar_nj}
\end{equation}
with the eigenenergies $\bar{E}_{n}(t)=-E_{n}(t)$. For the system $A$
with a bounded path, the geometric function $G_{n,m}^{p,q}(\vartheta)$
evolves finitely along the path. In Appendix~\ref{sec:MZ}, we obtain
\begin{equation}
\bar{G}_{n,m}^{p,q}(\vartheta)=\bar{F}_{n,m}^{*}(\vartheta)G_{n,m}^{p,q}(\vartheta),\label{eq:GnmpqBarVarTheta}
\end{equation}
which contains the fast oscillating factors $\bar{F}_{n,m}^{*}(\vartheta)=e^{i\int_{0}^{t}[E_{n}-E_{m}]dt^{\prime}}$.
Therefore in the adiabatic limit, the change of the geometric function $\bar{G}_{n,m}^{p,q}(\vartheta)$
is not finite and the path of $\bar{A}$ is not bounded.
The nonadiabatic evolution 
\begin{equation}\bar{U}_{\text{Dia}}(t)=\mathcal{P}\exp\left[i\int_{\vartheta_{0}}^{\vartheta}\sum_{n\neq m;p,q}G_{n,m}^{p,q}(\vartheta^{\prime})d\vartheta^{\prime}\right]
\end{equation}
becomes purely geometric and the effect of nonadiabatic evolution $\bar{U}_{\text{Dia}}(t)$ of the system $\bar{A}$ does not vanish for general paths.
Therefore the condition (\ref{eq:ADC1}) does not grantee finite eigenpaths
and is not sufficient. 

It was claimed that the condition (\ref{eq:ADC1})
is necessary when there is no energy degeneracy or crossings~\cite{Tong:2010:120401}.
We have shown that energy crossings are possible in the adiabatic
evolution. Thus the condition (\ref{eq:ADC1}) is also not necessary.
To have adiabatic evolution, the geometric operator $G_{n,m}^{p,q}(\vartheta)$
should be slowly varying compared with $F_{n,m}(\vartheta)$.

\section{Conclusions and discussions}
We have developed a gauge invariant formalism to obtain the whole
nonadiabatic transitions in the adiabatic approximation, and have used this to show that the instantaneous eigenenergies and eigenstates play different
roles in the adiabatic evolution. For finite evolution paths, the
instantaneous eigenenergies can change rapidly as long as the gap
modulations are off-resonant to the excitations generated by the instantaneous
eigenstates. We have demonstrated examples of adiabatic evolution by fast changing fields, which can lead to better adiabatic evolution. An arbitrary
number of level crossings during the adiabatic evolution is possible.
Under an exact and transparent formalism, we have shown by general
Hamiltonians with possible degeneracy and crossings that the Marzlin-Sanders
inconsistency arises because the evolution path is not slowly varying.
Our formalism also clearly show that the quantitative condition Eq.~(\ref{eq:ADC1})
is neither necessary nor sufficient. A necessary and sufficient condition
for adiabatic evolution has been provided.

Note that we can achieve $U_{\text{adia}}(t)$
by using the Hamiltonian $H^{\prime}(t)=i\dot{U}_{\text{adia}}(t)U_{\text{adia}}^{\dagger}(t)$, a scheme called transitionless or
counterdiabatic quantum driving~\cite{demirplak2003adiabatic,demirplak2008consistency,Berry:2009:365303,delCampo:2013:100502}. 
Since generally $|n_{j}^{t}\rangle$ is not the eigenstate of the driving Hamiltonian $H^{\prime}(t)$,
this driving does not follow
the adiabatic evolution.

\begin{acknowledgments}
Wang is grateful to Ren-Bao Liu for support and discussions, and thanks
Fan Yang and Alexander Crosse for discussions. The work is supported by Alexander von Humboldt Professorship and the DFG via SPP 1601. Wang thanks supports from the Hong Kong GRF CUHK402209, the CUHK Focused Investments Scheme, and the National Natural Science Foundation of China Project No. 11028510. We thank Tianyu Xie for careful reading.
\end{acknowledgments}

\appendix
\section{Gauge invariance\label{sec:GaugeInv}}

Consider the gauge transformation
\begin{equation}
|n_{j}^{t}\rangle\rightarrow|\tilde{n}_{j}^{t}\rangle=W_{t}|n_{j}^{t}\rangle,
\end{equation}
where the time-dependent unitary operator $W_{t}$ is the transformation
within each degenerate subspace with the property
\begin{equation}
\langle m_{p}^{t}|W_{t}|n_{q}^{t}\rangle=0,\text{ if }m\neq n.\label{eq:Wn}
\end{equation}
An example of this transformation is the phase shifts $W_{t}=\sum_{n,j} \exp(i\phi_{n,j}^{t})|n_{j}^{t}\rangle\langle n_{j}^{t}|$
of the eigenstates. The property Eq.~(\ref{eq:Wn}) leads to
\begin{equation}
\left[\sum_{j}|n_{j}^{t}\rangle\langle n_{j}^{t}|,W_{t}\right]=0,\label{eq:WtCommute}
\end{equation}
which can be verified by using Eq.~(\ref{eq:Wn}) and inserting the identity operator $I=\sum_{m,k}|m_{k}^{t}\rangle\langle m_{k}^{t}|$
into the commutator:

\begin{equation}
\sum_{j}|n_{j}^{t}\rangle\langle n_{j}^{t}|W_{t}\sum_{m,k}|m_{k}^{t}\rangle\langle m_{k}^{t}|-\sum_{m,k}|m_{k}^{t}\rangle\langle m_{k}^{t}|W_{t}\sum_{j}|n_{j}^{t}\rangle\langle n_{j}^{t}|=0.
\end{equation}

Using Eq.~(\ref{eq:WtCommute}), the system Hamiltonian
\begin{align}
\tilde{H}(t) & =\sum_{n,j}E_{n}(t)|\tilde{n}_{j}^{t}\rangle\langle\tilde{n}_{j}^{t}|=\sum_{n}E_{n}W_{t}\sum_{j}|n_{j}^{t}\rangle\langle n_{j}^{t}|W_{t}^{\dagger},\\
 & =\sum_{n,j}E_{n}(t)|n_{j}^{t}\rangle\langle n_{j}^{t}|=H(t),
\end{align}
is gauge invariant under the transformation of $W_{t}$.

\subsection{Gauge invariance of $U_{\text{Dyn}}(t)$}

Using Eq.~(\ref{eq:WtCommute}), the dynamic phase operator
\begin{align}
\tilde{U}_{\text{Dyn}}(t) & =\sum_{n,}e^{-i\int_{0}^{t}E_{n}(\tau)d\tau}\sum_{j}|\tilde{n}_{j}^{t}\rangle\langle\tilde{n}_{j}^{t}|,\\
 & =\sum_{n}e^{-i\int_{0}^{t}E_{n}(\tau)d\tau}\sum_{j}W_{t}|n_{j}^{t}\rangle\langle n_{j}^{t}|W_{t}^{\dagger},\\
 & =\sum_{n}e^{-i\int_{0}^{t}E_{n}(\tau)d\tau}\sum_{j}|n_{j}^{t}\rangle\langle n_{j}^{t}|=U_{\text{Dyn}}(t),
\end{align}
is gauge invariant.

\subsection{Gauge invariance of $U_{\text{Geo}}(t)$}

We first find the Hamiltonian $H_{W}(t)=i\dot{U}_{W}(t)U_{W}^{\dagger}(t)$
of the propagator
\begin{align}
U_{W}(t) & \equiv\left(\sum_{n,p}|n_{p}^{0}\rangle\langle n_{p}^{t}|\right)\left(\sum_{m,q}|\tilde{m}_{q}^{t}\rangle\langle\tilde{m}_{q}^{0}|\right),\\
 & =\sum_{n,p,q}|n_{p}^{0}\rangle\langle n_{p}^{t}|W_{t}|n_{q}^{t}\rangle\langle n_{q}^{0}|W_{0}^{\dagger},
\end{align}
where we have used Eq.~(\ref{eq:Wn}). We have $U_{W}(0)=I$ and
$U_{W}(t)=\mathcal{T}e^{-i\int_{0}^{t}H_{W}(s)ds}$. Using Eqs.~(\ref{eq:Wn})
and (\ref{eq:WtCommute}), we obtain
\begin{align}
H_{W}(t) & =i\sum_{n,p,q}|n_{p}^{0}\rangle\langle\dot{n}_{p}^{t}|n_{q}^{t}\rangle\langle n_{q}^{0}|+i\sum_{n,p,q}|n_{p}^{0}\rangle\langle n_{p}^{t}|\dot{W}_{t}W_{t}^{\dagger}|n_{q}^{t}\rangle\langle n_{q}^{0}|\nonumber \\
 & +i\sum_{n,p,q,j}|n_{p}^{0}\rangle\langle n_{p}^{t}|W_{t}|\dot{n}_{q}^{t}\rangle\langle n_{q}^{t}|W_{t}^{\dagger}|n_{j}^{t}\rangle\langle n_{j}^{0}|.\label{eq:Hw}
\end{align}

The geometric phase factor
\begin{equation}
\tilde{U}_{\text{Geo}}(t)=\tilde{U}_{\text{G1}}(t)\tilde{U}_{\text{G2}}(t),\label{eq:GaugeUGeo}
\end{equation}
where
\begin{align}
\tilde{U}_{\text{G1}}(t) & =\sum_{n,j}|\tilde{n}_{j}^{t}\rangle\langle\tilde{n}_{j}^{0}|=W_{t}\sum_{n,j}|n_{j}^{t}\rangle\langle n_{j}^{0}|W_{0}^{\dagger},\\
 & =U_{\text{G1}}(t)U_{W}(t),\label{eq:GaugeUG1}
\end{align}
and
\begin{equation}
\tilde{U}_{\text{G2}}(t)=\mathcal{T}\exp\left[-\int_{0}^{t}\sum_{n,p,q}|\tilde{n}_{p}^{0}\rangle\langle\tilde{n}_{p}^{t^{\prime}}|\dot{\tilde{n}}_{q}^{t^{\prime}}\rangle\langle\tilde{n}_{q}^{0}|dt^{\prime}\right].
\end{equation}
We rewrite $\tilde{U}_{\text{G2}}(t)$ as
\begin{equation}
\tilde{U}_{\text{G2}}(t)=U_{W}^{\dagger}(t)\mathcal{T}\exp\left[-i\int_{0}^{t}H_{WG2}(t^{\prime})dt^{\prime}\right],\label{eq:GaugeUwUG2}
\end{equation}
where the Hamiltonian
\begin{equation}
H_{WG2}(t)\equiv-iU_{W}(t)\sum_{n,p,q}|\tilde{n}_{p}^{0}\rangle\langle\tilde{n}_{p}^{t^{\prime}}|\dot{\tilde{n}}_{q}^{t^{\prime}}\rangle\langle\tilde{n}_{q}^{0}|U_{W}^{\dagger}(t)+H_{W}(t).
\end{equation}

Using Eqs.~(\ref{eq:Wn}) and (\ref{eq:WtCommute}), we obtain
\begin{align}
H_{WG2}(t) & =-i\sum_{n,p,q}|n_{p}^{0}\rangle\langle n_{p}^{t}|\dot{W}_{t}W_{t}^{\dagger}|n_{q}^{t}\rangle\langle n_{q}^{0}|+H_{W}(t)\nonumber \\
 & -i\sum_{n,p,q,j}|n_{p}^{0}\rangle\langle n_{p}^{t}|W_{t}|\dot{n}_{q}^{t}\rangle\langle n_{q}^{t}|W_{t}^{\dagger}|n_{j}^{t}\rangle\langle n_{j}^{0}|.\label{eq:HWG2}
\end{align}
Substituting Eq.~(\ref{eq:Hw}) into Eq.~(\ref{eq:HWG2}), we have
\begin{equation}
H_{WG2}(t)=i\sum_{n,p,q}|n_{p}^{0}\rangle\langle\dot{n}_{p}^{t}|n_{q}^{t}\rangle\langle n_{q}^{0}|=H_{\text{G2}}(t).\label{eq:HWG2final}
\end{equation}
From Eqs.~(\ref{eq:GaugeUGeo}), (\ref{eq:GaugeUG1}), (\ref{eq:GaugeUwUG2}),
and (\ref{eq:HWG2final}), we can see that
\begin{equation}
\tilde{U}_{\text{Geo}}(t)=U_{\text{G1}}(t)U_{\text{G2}}(t)=U_{\text{Geo}}(t)
\end{equation}
is gauge invariant.

\subsection{Gauge invariance of $U_{\text{Dia}}(t)$}

As $U_{\text{Geo}}(t)$ is gauge invariant, we just need to show
\begin{equation}
\sum_{p,q}|n_{p}^{t}\rangle\langle n_{p}^{t}|\dot{m}_{q}^{t}\rangle\langle m_{q}^{t}|,\text{ for }n\neq m,
\end{equation}
is gauge invariant. For $n\neq m$,
\begin{align}
\sum_{p,q}|\tilde{n}_{p}^{t}\rangle\langle\tilde{n}_{p}^{t}|\dot{\tilde{m}}_{q}^{t}\rangle&\langle\tilde{m}_{q}^{t}|=\sum_{p,q}W_{t}|n_{p}^{t}\rangle\nonumber \\
 & \times\left(\langle n_{p}^{t}|W_{t}^{\dagger}\dot{W}_{t}|m_{q}^{t}\rangle+\langle n_{p}^{t}|\dot{m}_{q}^{t}\rangle\right)\langle m_{q}^{t}|W_{t}^{\dagger}.
\end{align}
Using Eq.~(\ref{eq:WtCommute}), we have
\begin{align}
\sum_{p,q}|\tilde{n}_{p}^{t}\rangle\langle\tilde{n}_{p}^{t}|\dot{\tilde{m}}_{q}^{t}\rangle\langle\tilde{m}_{q}^{t}| & =\sum_{p,q}|n_{p}^{t}\rangle\langle n_{p}^{t}|\dot{W}_{t}|m_{q}^{t}\rangle\langle m_{q}^{t}|W_{t}^{\dagger}\nonumber \\
 & +\sum_{p,q}|n_{p}^{t}\rangle\langle n_{p}^{t}|W_{t}|\dot{m}_{q}^{t}\rangle\langle m_{q}^{t}|W_{t}^{\dagger}.\label{eq:npnpmqmq}
\end{align}
The time derivative of Eq.~(\ref{eq:Wn}) gives
\begin{equation}
\langle n_{p}^{t}|\dot{W}_{t}|m_{q}^{t}\rangle=-\langle\dot{n}_{p}^{t}|W_{t}|m_{q}^{t}\rangle-\langle n_{p}^{t}|W_{t}|\dot{m}_{q}^{t}\rangle,\text{ for }n\neq m.\label{eq:WdotReplacement}
\end{equation}
By substitution of Eq.~(\ref{eq:WdotReplacement}) into Eq.~(\ref{eq:npnpmqmq}),
we get
\begin{align}
\sum_{p,q}|\tilde{n}_{p}^{t}\rangle\langle\tilde{n}_{p}^{t}|\dot{\tilde{m}}_{q}^{t}\rangle\langle\tilde{m}_{q}^{t}| & =-\sum_{p,q}|n_{p}^{t}\rangle\langle\dot{n}_{p}^{t}|W_{t}|m_{q}^{t}\rangle\langle m_{q}^{t}|W_{t}^{\dagger}.
\end{align}
Using Eq.~(\ref{eq:WtCommute}) and $\frac{d}{dt}\left(\langle n_{p}^{t}|m_{q}^{t}\rangle\right)=\langle\dot{n}_{p}^{t}|m_{q}^{t}\rangle+\langle n_{p}^{t}|\dot{m}_{q}^{t}\rangle=0$,
we obtain for $n\neq m$,
\begin{align}
\sum_{p,q}|\tilde{n}_{p}^{t}\rangle\langle\tilde{n}_{p}^{t}|\dot{\tilde{m}}_{q}^{t}\rangle\langle\tilde{m}_{q}^{t}| & =\sum_{p,q}|n_{p}^{t}\rangle\langle n_{p}^{t}|\dot{m}_{q}^{t}\rangle\langle m_{q}^{t}|.
\end{align}
Therefore $U_{\text{Dia}}(t)$ is gauge invariant.

The gauge invariance of $U_{\text{Dia}}(t)$ can also be verified by the facts that $U_{\text{Dia}}(t)=[U_{\text{Dyn}}(t)U_{\text{Geo}}(t)]^{\dag}U(t)$ and $U_{\text{Dyn}}(t)$, $U_{\text{Geo}}(t)$, and $U(t)$ are gauge invariant.

\section{The proofs of necessity and sufficiency\label{sec:Condition}}

\subsection{Sufficiency}

For simplicity, we define $F_{\mu}\equiv F_{n,m}$ and $G_{\mu}\equiv\sum_{p,q}G_{n,m}^{p,q}$
in $U_{\text{Dia}}(t)$ and write it as $U_{\text{Dia}}(\vartheta)\equiv U_{\text{Dia}}(t)=\mathcal{P}\exp\left[i\int_{\vartheta_{0}}^{\vartheta}\sum_{\mu}F_{\mu}(\vartheta^{\prime})G_{\mu}(\vartheta^{\prime})d\vartheta^{\prime}\right]$
by using $\mu$ to indicate the summation over $n\neq m$. The nonadiabatic
deviation Eq.~(\ref{eq:Derr}) reads
\begin{equation}
D_{\text{Dia}}(t)=i\int_{\vartheta_{0}}^{\vartheta}\sum_{\mu}F_{\mu}G_{\mu}U_{\text{Dia}}d\vartheta^{\prime}.\label{eq:bound:Derr}
\end{equation}

We use a partition for the interval $[\vartheta_{0},\vartheta]$ by
$N-1$ points $\vartheta_{j}$, such that $\vartheta_{0}<\vartheta_{1}<\vartheta_{2}<\cdots<\vartheta_{N-1}<\vartheta\equiv\vartheta_{N}$
with the interval
\begin{equation}
\eta_{\text{min}}\leq\vartheta_{j+1}-\vartheta_{j}\leq\eta,
\end{equation}
for all $j=0,1,\cdots,N-1$. 
Let
\begin{equation}
g_{\text{tot}}\equiv\sum_{\mu}g_{\mu},
\end{equation}
with the least upper bound of the unitarily invariant norm
\begin{equation}
g_{\mu}\equiv\sup_{\vartheta^{\prime}\in[\vartheta_{0},\vartheta]}||G_{\mu}(\vartheta^{\prime})||.
\end{equation}
The change of $G_{\mu}(\vartheta^{\prime})$ is continuous, with
a finite time derivative for $\vartheta^{\prime}\in[\vartheta_{0},\vartheta]$,
and we define
\begin{subequations}
\begin{align}
w_{\text{tot}}&\equiv\sum_{\mu}w_{\mu},\\
w_{\mu}&\equiv\sup_{\vartheta^{\prime}\in[\vartheta_{0},\vartheta]}||\frac{d}{d\vartheta^{\prime}}G_{\mu}(\vartheta^{\prime})||.
\end{align}
\end{subequations}
Any bounded operator $A(\vartheta^{\prime})$ has an associate step
function $\overline{A}(\vartheta^{\prime})=A(\vartheta_{j})$ when
$\vartheta^{\prime}\in[\vartheta_{j},\vartheta_{j+1})$. For $\vartheta^{\prime}\in[\vartheta_{j},\vartheta_{j+1})$,
the difference
\begin{align}
\left\Vert G_{\mu}(\vartheta^{\prime})-\overline{G}_{\mu}(\vartheta^{\prime})\right\Vert  & =\left\Vert G_{\mu}(\vartheta^{\prime})-G_{\mu}(\vartheta_{j})\right\Vert,\\
 & =\left\Vert \int_{\vartheta_{j}}^{\vartheta^{\prime}}\frac{d}{d\theta}G_{\mu}(\theta)d\theta\right\Vert,\\
 & \leq(\vartheta^{\prime}-\vartheta_{j})w_{\mu},\\
 & <\eta w_{\mu}.\label{eq:bound:G-G}
\end{align}
For $\vartheta^{\prime}\in[\vartheta_{j},\vartheta_{j+1})$, the difference
\begin{align}
\left\Vert U_{\text{Dia}}(\vartheta^{\prime})-\overline{U}_{\text{Dia}}(\vartheta^{\prime})\right\Vert  & =\left\Vert U_{\text{Dia}}(\vartheta^{\prime})-U_{\text{Dia}}(\vartheta_{j})\right\Vert,\\
 & =\left\Vert \int_{\vartheta_{j}}^{\vartheta^{\prime}}\frac{d}{d\theta}U_{\text{Dia}}(\theta)d\theta\right\Vert,\\
 & \leq\sum_{\mu}\left\Vert \int_{\vartheta_{j}}^{\vartheta^{\prime}}F_{\mu}(\theta)G_{\mu}(\theta)U_{\text{Dia}}(\theta)d\theta\right\Vert,\\
 & \leq(\vartheta^{\prime}-\vartheta_{j})g_{\text{tot}},\\
 & <\eta g_{\text{tot}},\label{eq:bound:U-U}
\end{align}
where we have used $|F_{\mu}(\vartheta^{\prime})|=1$. From Eqs.~(\ref{eq:bound:G-G})
and (\ref{eq:bound:U-U}), we have the norm

\begin{gather}
\left\Vert G_{\mu}U_{\text{Dia}}-\overline{G}_{\mu}\overline{U}_{\text{Dia}}\right\Vert =\left\Vert G_{\mu}(U_{\text{Dia}}-\overline{U}_{\text{Dia}})+(G_{\mu}-\overline{G}_{\mu})\overline{U}_{\text{Dia}}\right\Vert, \nonumber \\
<\eta\left(g_{\mu}g_{\text{tot}}+w_{\mu}\right).
\end{gather}

We write the deviation Eq.~(\ref{eq:bound:Derr}) as $D_{\text{Dia}}(\vartheta)\equiv D_{\text{Dia}}^{(1)}+D_{\text{Dia}}^{(2)}$,
where the error caused by the partition
\begin{equation}
D_{\text{Dia}}^{(1)}=i\int_{\vartheta_{0}}^{\vartheta}\sum_{\mu}F_{\mu}\left[G_{\mu}U_{\text{Dia}}-\overline{G}_{\mu}\overline{U}_{\text{Dia}}\right]d\vartheta^{\prime}
\end{equation}
has the norm

\begin{align}
\left\Vert D_{\text{Dia}}^{(1)}\right\Vert  & <\eta\left(g_{\text{tot}}^{2}+w_{\text{tot}}\right)(\vartheta-\vartheta_{0}),\label{eq:bound:b1}
\end{align}
and
\begin{equation}
D_{\text{Dia}}^{(2)}=i\int_{\vartheta_{0}}^{\vartheta}\sum_{\mu}F_{\mu}\overline{G}_{\mu}\overline{U}_{\text{Dia}}d\vartheta^{\prime}.
\end{equation}
Under the averaging condition \ref{eq:averagingCondition}
\begin{equation}
\left|\int_{\vartheta_{0}}^{\vartheta^{\prime}}F_{n,m}(\vartheta^{\prime})d\vartheta^{\prime}\right|<\xi_{\text{avg}},\text{ for }\vartheta^{\prime}\in[\vartheta_{0},\vartheta]\text{ and }n\neq m,\label{eq:bound:avg}
\end{equation}
we have the norm
\begin{align}
\left\Vert D_{\text{Dia}}^{(2)}\right\Vert  & =\left\Vert \sum_{\mu}\sum_{j}\overline{G}_{\mu}\overline{U}_{\text{Dia}}\int_{\vartheta_{j}}^{\vartheta_{j+1}}F_{\mu}d\vartheta\right\Vert, \\
 & \leq\sum_{\mu}\sum_{j}\left\Vert \overline{G}_{\mu}\overline{U}_{\text{Dia}}\int_{\vartheta_{j}}^{\vartheta_{j+1}}F_{\mu}d\vartheta\right\Vert, \\
 & <\xi_{\text{avg}}Ng_{\text{tot}}.\label{eq:bound:b2}
\end{align}
The nonadiabatic deviation 
\begin{equation}
\left\Vert D_{\text{Dia}}\right\Vert \leq\left\Vert D_{\text{Dia}}^{(1)}\right\Vert +\left\Vert D_{\text{Dia}}^{(2)}\right\Vert. \label{eq:bound:b12}
\end{equation}

For sufficiently small $\xi_{\text{avg}}\ll (\vartheta-\vartheta_{0})^{2}\left(g_{\text{tot}}^{2}+w_{\text{tot}}\right)/g_{\text{tot}}$, we choose the partition with $\eta\approx\eta_{\text{min}}\approx \sqrt{\xi_{\text{avg}}g_{\text{tot}}/\left(g_{\text{tot}}^{2}+w_{\text{tot}}\right)}\ll(\vartheta-\vartheta_{0})$. With this partition, we obtain $\left\Vert D_{\text{Dia}}^{(1)}\right\Vert \approx \left\Vert D_{\text{Dia}}^{(2)}\right\Vert $ and the upper bound Eq.~(\ref{eq:bound:type1}).

We may choose other partitions to obtain other bounds. For example, for $\xi_{\text{avg}}\ll 1$, we choose $\eta_{\text{min}}=\eta=(\vartheta-\vartheta_{0})/N$ with $N$ the smallest integer greater than $1/\sqrt{\xi_{\text{avg}}}$. We have for this partition 
\begin{align}
\eta & <\sqrt{\xi_{\text{avg}}}(\vartheta-\vartheta_{0}), \label{eq:etaLess} \\
N & \leq (1/\sqrt{\xi_{\text{avg}}})+1. \label{eq:NLess}
\end{align}
Using Eqs.~(\ref{eq:bound:b1}), (\ref{eq:bound:b2}), (\ref{eq:bound:b12}), (\ref{eq:etaLess}), and (\ref{eq:NLess}), we obtain the upper bound Eq.~(\ref{eq:bound:type2}) for $\xi_{\text{avg}}\ll 1$.

Therefore \begin{equation}\lim_{\xi_{\text{avg}}\rightarrow0}U_{\text{Dia}}(t)=I.
\end{equation}
and the averaging condition \ref{eq:averagingCondition} with $\xi_{\text{avg}}\ll 1$ is sufficient. 

\subsubsection*{Derivation of Eq.~(\ref{eq:DBound})}
For the partition that the average
\begin{equation}
\int_{\vartheta_{j}}^{\vartheta_{j+1}}F_{n,m}(\vartheta^{\prime})d\vartheta^{\prime}=0,\text{ for }n\neq m,
\end{equation}
vanishes for all the intervals $j=0,1,\ldots,N-1$,
we have $ D_{\text{Dia}}^{(2)} = 0$, $\xi_{\text{avg}}=\eta$, and
\begin{equation}
\left\Vert D_{\text{Dia}}(t)\right\Vert <\xi_{\text{avg}}\left(g_{\text{tot}}^{2}+w_{\text{tot}}\right)(\vartheta-\vartheta_{0}),
\end{equation}
which is simpler than the bounds Eqs.~(\ref{eq:bound:type1}) and (\ref{eq:bound:type2}).

\subsection{Necessity}

A general condition for adiabatic evolution should be universal and works for all
bounded paths. We choose a path that satisfies $\frac{d}{dt}|n_{p}^{\vartheta}\rangle=0$
if $n_{p}\neq N,M$ and the states $|N^{\vartheta}\rangle=\cos(b\vartheta)|N^{\vartheta_{0}}\rangle-i\sin(b\vartheta)|M^{\vartheta_{0}}\rangle$
and $|M^{\vartheta}\rangle=-i\sin(b\vartheta)|N^{\vartheta_{0}}\rangle+\cos(b\vartheta)|M^{\vartheta_{0}}\rangle$
with $b=O(1)$. We have $|\frac{d}{d\vartheta}N^{\vartheta}\rangle=-ib|M^{\vartheta}\rangle$,
$|\frac{d}{d\vartheta}M^{\vartheta}\rangle=-ib|N^{\vartheta}\rangle$,
and thus $U_{\text{G2}}(\vartheta)=I$ by using Eq.~(\ref{eq:UG2}).
The deviation from the adiabatic evolution is

\begin{gather}
D_{\text{Dia}}(\vartheta)=i\int_{\vartheta_{0}}^{\vartheta}b\left[F_{N,M}(\vartheta^{\prime})|N^{\vartheta_{0}}\rangle\langle M^{\vartheta_{0}}|+\text{H.c.}\right]U_{\text{Dia}}(\vartheta^{\prime})d\vartheta^{\prime}.
\end{gather}
Using $U_{\text{Dia}}(\vartheta^{\prime})=D_{\text{Dia}}(\vartheta^{\prime})+I$,
we write
\begin{gather}
\int_{\vartheta_{0}}^{\vartheta}b\left[F_{N,M}(\vartheta^{\prime})|N^{\vartheta_{0}}\rangle\langle M^{\vartheta_{0}}|+\text{H.c.}\right]d\vartheta^{\prime}=\nonumber \\
iD_{\text{Dia}}(\vartheta)-\int_{\vartheta_{0}}^{\vartheta}b\left[F_{N,M}(\vartheta^{\prime})|N^{\vartheta_{0}}\rangle\langle M^{\vartheta_{0}}|+\text{H.c.}\right]D_{\text{Dia}}(\vartheta^{\prime})d\vartheta^{\prime}.\label{eq:bound:DerrP}
\end{gather}
For a good adiabatic approximation, the correction $\left\Vert D_{\text{Dia}}(\vartheta^{\prime})\right\Vert <\epsilon$
is small for all bounded paths $\vartheta^{\prime}\in[\vartheta_{0},\vartheta]$. Here $\epsilon$ is a small value.
By choosing other paths with different $|N^{\vartheta_{0}}\rangle$
and $|M^{\vartheta_{0}}\rangle$ in Eq.~(\ref{eq:bound:DerrP}),
we have for $n\neq m$,
\begin{equation}
\left|\int_{\vartheta_{0}}^{\vartheta}F_{n,m}(\vartheta^{\prime})d\vartheta^{\prime}\right|<\epsilon\kappa,
\end{equation}
with a finite $\kappa$. In the adiabatic limit
\begin{equation}
\lim_{\left\Vert D_{\text{Dia}}\right\Vert \rightarrow0}\left|\int_{\vartheta_{0}}^{\vartheta}F_{n,m}(\vartheta^{\prime})d\vartheta^{\prime}\right|=0,
\end{equation}
for $n\neq m$.

\section{Derivation of Eq.~(\ref{eq:GnmpqBarVarTheta}) \label{sec:MZ}}
We express the geometric function for the system $\bar{A}$ ($n\neq m$)
\begin{align}
\bar{G}_{n,m}^{p,q}(\vartheta) & =\bar{U}_{\text{Geo}}^{\dagger}(\vartheta)|\bar{n}_{p}^{\vartheta}\rangle\left(\langle\bar{n}_{p}^{\vartheta}|i\frac{d}{d\vartheta}|\bar{m}_{q}^{\vartheta}\rangle\right)\langle\bar{m}_{q}^{\vartheta}|\bar{U}_{\text{Geo}}(\vartheta)
\end{align}
by the time parameter $t$ as
\begin{equation}
\bar{G}_{n,m}^{p,q}(t)=i\bar{U}_{\text{Geo}}^{\dagger}(t)|\bar{n}_{p}^{t}\rangle\langle\bar{n}_{p}^{t}|\dot{\bar{m}}_{q}^{t}\rangle\langle\bar{m}_{q}^{t}|\bar{U}_{\text{Geo}}(t)\frac{dt}{d\vartheta}.\label{eq:GnmpqBar}
\end{equation}
By using $\bar{U}_{\text{Geo}}(t)=\bar{U}_{\text{G1}}(t)\bar{U}_{\text{G2}}(t)$,
$\bar{U}_{\text{G1}}(t)=\sum_{n,j}|\bar{n}_{j}^{t}\rangle\langle\bar{n}_{j}^{0}|$,
and $\bar{U}_{\text{G2}}(t)=\mathcal{T}\exp[-i\int_{0}^{t}\bar{H}_{\text{G2}}(t^{\prime})dt^{\prime}]$
with $\bar{H}_{\text{G2}}(t)=-i\sum_{n,p,q}|\bar{n}_{p}^{0}\rangle\langle\bar{n}_{p}^{t}|\dot{\bar{n}}_{q}^{t}\rangle\langle\bar{n}_{q}^{0}|$
for the system $\bar{A}$, we get 
\begin{equation}
\bar{G}_{n,m}^{p,q}(t)=i\bar{U}_{\text{G2}}^{\dagger}(t)|\bar{n}_{p}^{0}\rangle\langle\bar{n}_{p}^{t}|\dot{\bar{m}}_{q}^{t}\rangle\langle\bar{m}_{q}^{0}|\bar{U}_{\text{G2}}(t)\frac{dt}{d\vartheta}.\label{eq:GnmpqBar2}
\end{equation}

To relate the expressions to the quantities of system $A$, we use Eq.~(\ref{eq:njBar_nj})
to obtain 
\begin{equation}
|\bar{n}_{j}^{0}\rangle=|n_{j}^{0}\rangle,\label{eq:njBarEQnj}
\end{equation}
and
\begin{align}
|\dot{\bar{m}}_{q}^{t}\rangle & =[\frac{d}{dt}U^{\dagger}(t)]|m_{q}^{t}\rangle+U^{\dagger}(t)|\dot{m}_{q}^{t}\rangle\\
 & =iU^{\dagger}(t)H(t)|m_{q}^{t}\rangle+U^{\dagger}(t)|\dot{m}_{q}^{t}\rangle\\
 & =iE_{m}(t)U^{\dagger}(t)|m_{q}^{t}\rangle+U^{\dagger}(t)|\dot{m}_{q}^{t}\rangle.\label{eq:mqtDot}
\end{align}
By using Eqs.~(\ref{eq:njBar_nj}), (\ref{eq:njBarEQnj}) and (\ref{eq:mqtDot}),
Eq.~(\ref{eq:GnmpqBar2}) becomes 
\begin{equation}
\bar{G}_{n,m}^{p,q}(t)=i\bar{U}_{\text{G2}}^{\dagger}(t)|n_{p}^{0}\rangle\langle n_{p}^{t}|\dot{m}_{q}^{t}\rangle\langle m_{q}^{0}|\bar{U}_{\text{G2}}(t)\frac{dt}{d\vartheta},\label{eq:GnmpqBar3}
\end{equation}
for $n\neq m$. Similarly,
\begin{equation}
\bar{H}_{\text{G2}}(t)=-i\sum_{n,p,q}|n_{p}^{0}\rangle\langle n_{p}^{t}|\dot{n}_{q}^{t}\rangle\langle n_{q}^{0}|+\sum_{n,j}E_{n}(t)|n_{j}^{0}\rangle\langle n_{j}^{0}|.
\end{equation}
and hence
\begin{equation}
\bar{U}_{\text{G2}}(t)=\sum_{n,j}e^{-i\int_{0}^{t}E_{n}dt^{\prime}}|n_{j}^{0}\rangle\langle n_{j}^{0}|U_{\text{G2}}(t).\label{eq:UG2Bar}
\end{equation}
Substituting Eq.~(\ref{eq:UG2Bar}) into (\ref{eq:GnmpqBar3}) and using 
the geometric function for the system $A$ [cf. Eq.~(\ref{eq:GnmpqBar2})],
\begin{equation}
G_{n,m}^{p,q}(t)=iU_{\text{G2}}^{\dagger}(t)|n_{p}^{0}\rangle\langle n_{p}^{t}|\dot{m}_{q}^{t}\rangle\langle m_{q}^{0}|U_{\text{G2}}(t)\frac{dt}{d\vartheta},
\end{equation}
we obtain
\begin{align}
\bar{G}_{n,m}^{p,q}(t) & =e^{i\int_{0}^{t}(E_{n}-E_{m})dt^{\prime}}G_{n,m}^{p,q}(t)\\ 
 & =F_{n,m}(t)G_{n,m}^{p,q}(t)=\bar{F}_{n,m}^{*}(t)G_{n,m}^{p,q}(t),
\end{align}
which is Eq.~(\ref{eq:GnmpqBarVarTheta}).


\begin{thebibliography}{10}
\providecommand{\url}[1]{#1}
\csname url@samestyle\endcsname
\providecommand{\newblock}{\relax}
\providecommand{\bibinfo}[2]{#2}
\providecommand{\BIBentrySTDinterwordspacing}{\spaceskip=0pt\relax}
\providecommand{\BIBentryALTinterwordstretchfactor}{4}
\providecommand{\BIBentryALTinterwordspacing}{\spaceskip=\fontdimen2\font plus
\BIBentryALTinterwordstretchfactor\fontdimen3\font minus
  \fontdimen4\font\relax}
\providecommand{\BIBforeignlanguage}[2]{{%
\expandafter\ifx\csname l@#1\endcsname\relax
\typeout{** WARNING: IEEEtran.bst: No hyphenation pattern has been}%
\typeout{** loaded for the language `#1'. Using the pattern for}%
\typeout{** the default language instead.}%
\else
\language=\csname l@#1\endcsname
\fi
#2}}
\providecommand{\BIBdecl}{\relax}
\BIBdecl

\bibitem{Messiah:1965:North}
A.~Messiah, \emph{Quantum Mechanics, Vol. II}.\hskip 1em plus 0.5em minus
  0.4em\relax Amsterdam: North-Holland Publishing Co., 1965.

\bibitem{Berry:1984:45}
M. V.~Berry, ``Quantal phase factors accompanying adiabatic changes,'' \emph{Proc.
  Roy. Soc. London Ser. A}, vol. 392, p.~45, 1984.

\bibitem{Wilczek:1984:2111}
F.~Wilczek and A.~Zee, ``Appearance of gauge structure in simple dynamical
  systems,'' \emph{Phys. Rev. Lett.}, vol.~52, p. 2111, 1984.

\bibitem{Marzlin:2004:160408}
K.-P. Marzlin and B.~C. Sanders, ``Inconsistency in the application of the
  adiabatic theorem,'' \emph{Phys. Rev. Lett.}, vol.~93, p. 160408, 2004.

\bibitem{Tong:2005:110407}
D.~M. Tong, K.~Singh, L.~C. Kwek, and C.~H. Oh, ``Quantitative conditions do
  not guarantee the validity of the adiabatic approximation,'' \emph{Phys. Rev.
  Lett.}, vol.~95, p. 110407, 2005.

\bibitem{Sarandy:2004:331}
M. S.~Sarandy, L.-A. Wu, and D A.~Lidar, ``Consistency of the adiabatic theorem,''
  \emph{Quant. Inf. Proc.}, vol.~3, p. 331, 2004.

\bibitem{Du:2008:060403}
J.~Du, L.~Hu, Y.~Wang, J.~Wu, M.~Zhao, and D.~Suter, ``Experimental study of
  the validity of quantitative conditions in the quantum adiabatic theorem,''
  \emph{Phys. Rev. Lett.}, vol. 101, p. 060403, 2008.

\bibitem{Amin:2009:220401}
M.~H.~S. Amin, ``Consistency of the adiabatic theorem,'' \emph{Phys. Rev.
  Lett.}, vol. 102, p. 220401, 2009.

\bibitem{Tong:98:150402}
D.~M. Tong, K.~Singh, L.~C. Kwek, and C.~H. Oh, ``Sufficiency criterion for the
  validity of the adiabatic approximation,'' \emph{Phys. Rev. Lett.}, vol.~98,
  p. 150402, 2007.

\bibitem{Wei:2007:024304}
Z.~Wei and M.~Ying, ``Quantum adiabatic computation and adiabatic conditions,''
  \emph{Phys. Rev. A}, vol.~76, p. 024304, 2007.

\bibitem{Comparat:2009:012106}
D.~Comparat, ``General conditions for quantum adiabatic evolution,''
  \emph{Phys. Rev. A}, vol.~80, p. 012106, 2009.

\bibitem{lidar:2009:102106}
D.~A. Lidar, A.~T. Rezakhani, and A.~Hamma, ``Adiabatic approximation with
  exponential accuracy for many-body systems and quantum computation,''
  \emph{J. Math. Phys.}, vol.~50, p. 102106, 2009.

\bibitem{Boixo:2010:032308}
S.~Boixo and R.~D. Somma, ``Necessary condition for the quantum adiabatic
  approximation,'' \emph{Phys. Rev. A}, vol.~81, p. 032308, 2010.

\bibitem{Rigolin2012Adiabatic}
G.~Rigolin and G.~Ortiz, ``Adiabatic theorem for quantum systems with spectral
  degeneracy,'' \emph{Phys. Rev. A}, vol.~85, p. 062111, 2012.

\bibitem{Guo2013Nonperturbative}
C.~Guo, Q.-H. Duan, W.~Wu, and P.-X. Chen, ``Nonperturbative approach to the
  quantum adiabatic condition,'' \emph{Phys. Rev. A}, vol.~88, p. 012114, 2013.

\bibitem{Tong:2010:120401}
D.~M. Tong, ``Quantitative condition is necessary in guaranteeing the validity
  of the adiabatic approximation,'' \emph{Phys. Rev. Lett.}, vol. 104, p.
  120401, 2010.

\bibitem{Zhao2011Comment}
M.~Zhao and J.~Wu, ``Comment on “quantitative condition is necessary in
  guaranteeing the validity of the adiabatic approximation”,'' \emph{Phys.
  Rev. Lett.}, vol. 106, p. 138901, 2011.

\bibitem{Comparat2011Comment}
D.~Comparat, ``Comment on “quantitative condition is necessary in
  guaranteeing the validity of the adiabatic approximation”,'' \emph{Phys.
  Rev. Lett.}, vol. 106, p. 138902, 2011.

\bibitem{Tong2011Tong}
D.~M. Tong, ``Tong replies:,'' \emph{Phys. Rev. Lett.}, vol. 106, p. 138903, 2011.

\bibitem{Li:2014:053023}
D.~Li and M.-H. Yung, ``Why the quantitative condition fails to reveal quantum
  adiabaticity,'' \emph{New J. Phys.}, vol.~16, p. 053023,
  2014.

\bibitem{Avron:1999:203}
J.~E. Avron and A.~Elgart, ``Adiabatic theorem without a gap condition,''
  \emph{Commun. Math. Phys.}, vol. 203, p. 445, 1999.


\bibitem{Feynman:1951:108}
R.~P. Feynman, ``An operator calculus having applications in quantum
  electrodynamics,'' \emph{Phys. Rev.}, vol.~84, p. 108, 1951.

\bibitem{Lidar:2013:QEC}
D.~A. Lidar and T.~A. Brun, \emph{Quantum Error Correction}.\hskip 1em plus
  0.5em minus 0.4em\relax Cambridge University Press, 2013.

\bibitem{comment:operatorNorm}
{The results hold for any unitarily invariant norm, e.g., for the operator
  norm.}

\bibitem{Kahane:1980:108}
C.~S. Kahane, ``Generalizations of the Riemann-Lebesgue and Cantor-Lebesgue
  lemmas,'' \emph{Czech. Math. J.}, vol.~30, p. 108, 1980.

\bibitem{Li:2008:229}
Y.~T. Li and R.~Wong, ``Integral and series representations of the dirac delta
  function,'' \emph{Commun. Pure Appl. Anal.}, vol.~7, p. 229, 2008.

\bibitem{comment:H0}
{An example is a spin in a magnetic field $\boldsymbol{B}(t)$, and $H(t)=0$
  when $\boldsymbol{B}(t)=0$. For bare Hamiltonians that do not vanish, we may
  work in the rotating frame of the bare Hamiltonian and in this frame $H(t)=0$
  when there is no additional control field.}

\bibitem{Viola:1998:2733}
L.~Viola and S.Lloyd, ``Dynamical suppression of decoherence in two-state
  quantum systems,'' \emph{Phys. Rev. A}, vol.~58, p. 2733, 1998.

\bibitem{BanJMO1998}
M.~Ban, ``Photon-echo technique for reducing the decoherence of a quantum
  bit,'' \emph{J. Mod. Opt.}, vol.~45, p. 2315, 1998.

\bibitem{Yang:2010:2}
W.~Yang, Z.-Y. Wang, and R.-B. Liu, ``Preserving qubit coherence by dynamical
  decoupling,'' \emph{Front. Phys.}, vol.~6, p.~2, 2011.

\bibitem{Wang:2013:042319}
Z.-Y. Wang and R.-B. Liu, ``No-go theorems and optimization of dynamical
  decoupling against noise with soft cutoff,'' \emph{Phys. Rev. A}, vol.~87, p.
  042319, 2013.

\bibitem{john1996mathematical}
J.~von Neumann, \emph{Mathematical foundations of quantum mechanics}.\hskip 1em
  plus 0.5em minus 0.4em\relax Princeton university press, 1996.

\bibitem{Childs:2002:032314}
A.~M. Childs, E.~Deotto, E.~Farhi, J.~Goldstone, S.~Gutmann, and A.~J. Landahl,
  ``Quantum search by measurement,'' \emph{Phys. Rev. A}, vol.~66, p. 032314,
  2002.

\bibitem{Boixo:2009:833}
S.~Boixo, E.~H. Knill, and R.~Somma, ``Eigenpath traversal by phase
  randomization,'' \emph{Quant. Inf. \& Comp.}, vol.~9, p. 833, 2009.

\bibitem{Chiang:2014:012314}
H.-T. Chiang, G.~Xu, and R.~D. Somma, ``Improved bounds for eigenpath
  traversal,'' \emph{Phys. Rev. A}, vol.~89, p. 012314, 2014.





\bibitem{Carr:1954:630}
H.~Y. Carr and E.~M. Purcell, ``Effects of diffusion on free precession in
  nuclear magnetic resonance experiments,'' \emph{Phys. Rev.}, vol.~94, p. 630,
  1954.

\bibitem{Leek:2007:1889}
P.~J. Leek, J.~M. Fink, A.~Blais, R.~Bianchetti, M.~G\"oppl, J.~M. Gambetta,
  D.~I. Schuster, L.~Frunzio, R.~J. Schoelkopf, and A.~Wallraff, ``Observation
  of Berry's phase in a solid-state qubit,'' \emph{Science}, vol. 318, p. 1889,
  2007.

\bibitem{Berger:2012:220502}
S.~Berger, M.~Pechal, S.~Pugnetti, A.~A. Abdumalikov Jr., L.~Steffen, A.~Fedorov,
  A.~Wallraff, and S.~Filipp, ``Geometric phases in superconducting qubits
  beyond the two-level approximation,'' \emph{Phys. Rev. B}, vol.~85, p.
  220502, 2012.

\bibitem{Shevchenko:2010:1}
S. N.~Shevchenko, S.~Ashhab, and F.~Nori, ``Landau-Zener-St\"uckelberg
  interferometry,'' \emph{Phys. Rep.}, vol. 492, p.~1, 2010.
  

\bibitem{demirplak2003adiabatic}
M.~Demirplak and S.~A. Rice, ``Adiabatic population transfer with control
  fields,'' \emph{J. Phys. Chem. A}, vol. 107,  p. 9937, 2003.

\bibitem{demirplak2008consistency}
M.~Demirplak and S.~A. Rice, ``On the consistency, extremal, and global properties of
  counterdiabatic fields,'' \emph{J. Chem. Phys.}, vol. 129, p. 154111, 2008.

\bibitem{Berry:2009:365303}
M.~V. Berry, ``Transitionless quantum driving,'' \emph{J. Phys. A: Math.
  Theor.}, vol.~42, p. 365303, 2009.

\bibitem{delCampo:2013:100502}
A.~del Campo, ``Shortcuts to adiabaticity by counterdiabatic driving,''
  \emph{Phys. Rev. Lett.}, vol. 111, p. 100502, 2013.


\end{thebibliography}

\end{document}